# Topology Control and Network Lifetime in Three-Dimensional Wireless Sensor Networks

S. M. Nazrul Alam
Department of Computer Science
Cornell University
Ithaca NY 14853 USA
*smna@cs.cornell.edu*

Zygmunt J. Haas
Department of Electrical and Computer Engineering
Cornell University
Ithaca NY 14853 USA
*haas@ece.cornell.edu*

*http://wnl.ece.cornell.edu*

**ABSTRACT**

Coverage and connectivity issues of three-dimensional (3D) networks are addressed in [2], but that work assumes that a node can be placed at any arbitrary location. In this work, we drop that assumption and rather assume that nodes are uniformly and densely deployed in a 3D space. We want to devise a mechanism that keeps some nodes active and puts other nodes into sleep so that the number of active nodes at a time is minimized (and thus network life time is maximized), while maintaining full coverage and connectivity. One simple way to do that is to partition the 3D space into cells, and only one node in each cell remains active at a time. Our results show that the number of active nodes can be minimized if the shape of each cell is a truncated octahedron. It requires the sensing range to be at least 0.542326 times the transmission radius. This value is 0.5, 0.53452 and 0.5 for cube, hexagonal prism, and rhombic dodecahedron, respectively. However, at a time the number of active nodes for cube, hexagonal prism and rhombic dodecahedron model is respectively 2.372239, 1.82615 and 1.49468 times of that of truncated octahedron model. So clearly truncated octahedron model has the highest network lifetime. We also provide a distributed topology control algorithm that can be used by each sensor node to determine its cell id using a constant number of local arithmetic operations provided that the sensor node knows its location. We also validate our results by simulation.

## 1. INTRODUCTION

Most current wireless sensor networks research assume that the sensors are deployed on a two-dimensional (2D) plane. This is a good approximation for applications where sensors are deployed on earth surface and where the height of the network is smaller than transmission radius of a node. In these networks, the height of the network is negligible as compared to the length and the width. However, this 2D assumption is violated in underwater, atmospheric and space applications where height of the network can be significant and nodes are distributed over a three-dimensional (3D) space. Although such networks may not exist at present, there are works in progress that will make 3D networks increasingly common in the near future. For example, underwater ad hoc and sensor networks have attracted a lot of attention recently [1][8][10][13]. In underwater sensor networks, nodes may be placed at different depths of an ocean and thus the network becomes three-dimensional. Better weather forecasting and climate monitoring can be done by deploying three-dimensional networks in the atmosphere.

The issue of coverage and connectivity in 3D networks has been addressed in [2]. That work assumes that nodes can be placed anywhere with any arbitrary precision in a 3D space and solves the problem of finding optimal placement of nodes such that full coverage of a 3D space is achieved with minimum number of nodes. In this paper, we drop the assumption that nodes can be placed at any arbitrary place with arbitrary precision in a 3D space. Rather we assume that sensors are uniformly and densely distributed in the 3D space. Our topology control algorithm partitions the 3D space into equal sized cells. One sensor per cell is kept active for sensing and networking functions, while other sensors in the cell are kept asleep to increase network lifetime and to reduce traffic load (by decreasing redundant data). When the active node depletes its battery, another sensor node wakes up to become the active node. Similar approach has been explored in the context of 2D networks in [14][15].

In this paper, we want to answer the following questions:

- What is the best way to partition the network into cells in three-dimension such that for a fixed transmission radius the number of active nodes in the whole network is minimized while maintaining full connectivity?
- What should be the minimum sensing range in terms of transmission radius to ensure full coverage?
- How this partitioning can be done locally in a distributed way?

Reducing the number of active nodes directly contributes to the extension of network lifetime. However, maintaining full connectivity requires that the maximum distance between the active nodes of any two first-tier neighboring cells cannot exceed the transmission radius (i.e., communication range). Since active node can be located anywhere inside a cell, the maximum distance between two furthest points of two first-tier neighboring cells must be less than or equal to the transmission radius. The answer of second question is useful during the design phase of the sensor nodes. The answer to third question allows us to make the network scalable.

As mentioned in [2], proving optimality in many 3D problems is surprisingly difficult, even though proofs for similar problems in 2D can be found easily. For example, it took four hundred years to prove Kepler conjecture [7] and it is not known yet if Kelvin's 1887 conjecture is correct for identical cells (Kelvin conjecture has been shown to be incorrect when cells can have different shapes) [17][18]. So the approach of [2] is used here by comparing most likely shapes of the cell. Certainly, the shape of the cell must be a polyhedron that tessellates a 3D space. Cube is the only regular polyhedron that tessellates a 3D space. However, the performance of the network turns out to be very poor if the shape of the cell is a cube. The solution to our problem in 2D is the hexagon [14]. But the polyhedron that has hexagon as its cross section along all three axes in 3D does not have the space-filling property (i.e., it does not tessellate in 3D). The polyhedrons that have space-filling property as well as have at least one hexagonal cross-section are hexagonal prism and rhombic dodecahedron. Finally, the polyhedron that has the highest volume to radius ratio (i.e., volumetric quotient in [2]) among all the polyhedrons that tessellate a 3D space is truncated octahedron. We compare all of these four polyhedrons analytically and show that the answer to our first question is that the ideal shape of the cell is truncated octahedron.

We also provide a distributed topology control algorithm that partitions a 3D space into truncated octahedron shaped cells where each sensor can determine its location correctly. Localization in sensor network is an active area of research and accurate location determination is important for many sensor network applications. So we assume that there exists a localization component that allows a sensor node to determine its location accurately.

Our contributions, results and conclusions of this paper can be summarized as follows:

- Using the approach used in [2], we claim that the answer to our first question is to partition the 3D space such that each partition has the shape of a truncated octahedron. We compare it with other major polyhedrons that tessellate a 3D space, namely, cube, hexagonal prism and rhombic dodecahedron. Our results show that if the shape of the partition is cube (*CB*), then the number of active nodes required are $\frac{96\sqrt{3}}{17\sqrt{17}} = 2.372239$ times the number of active nodes required when the shape of a partition is a truncated octahedron (*TO*). For hexagonal prism (*HP*) and rhombic dodecahedron model (*RD*) model, this value is $\frac{28\sqrt{14}}{17\sqrt{17}} = 1.49468$ and $\frac{128}{17\sqrt{17}} = 1.82615$, respectively. Clearly, *TO* shaped partition requires significantly smaller number of active nodes.

- Since the nodes are uniformly distributed, the number of nodes in a cell is proportional to the volume of a cell. Our results show that the volume of a *CB* shaped cell is just 42.154% of that of *TO* shaped cell. The volume of an *HP* cell is 66.9% of that of a *TO* cell, and an RD cell has just 54.76% volume of a *TO* cell. Clearly, in each cell *TO* model has significantly larger number of sensor nodes to take over as the active node than other models which in turn means that network lifetime is significantly higher in *TO* model as compared to other models.

- However, the extra benefit of *TO* shaped cells comes with the price in terms of sensing range. *TO* model requires the sensing range to be at least 0.542326 times the transmission range. *HP* model requires sensing range to be at least 0.53452 times the transmission range and both *CB* and *RD* model requires this value to be 0.5.

- We provide a simple topology control algorithm that partitions the whole 3D network into truncated octahedral shaped virtual cells in a distributed fashion. Initially, the information sink broadcast a message containing its location. Once a sensor node knows its own location, location of the information sink and transmission radius, it can easily calculate in which cell it belongs to using equation (1) and (2).

The rest of the paper is organized as follows. Section 2 presents some relevant background information. Section 3 formally introduces the problem and the assumptions used. In section 4 we analyze the problem and derive the results. Simulation results are provided in Section 5. We discuss future works in Section 6 and the paper is concluded in Section 7.

## 2. BACKGROUND

Detail explanation on different kind of polyhedrons and other necessary background information on three-dimensional networks are provided in [2]. Although those background materials are very relevant to this work, due to lack of space we choose not to repeat them here. The reader is strongly encouraged to have a look at [2].

Extensive research has been done on topology control and network lifetime issues in 2D wireless ad hoc and sensor networks [3][12][14] [19] [20][21]. However, we are not aware of any significant work on these issues in the context of 3D network. In fact, it is hard to find references on three-dimensional networks in general. In addition to [2], we found only two works in the literature of cellular mobile networks [4][5]. Both studied 3D cellular networks; each cell is represented as rhombic dodecahedron in [4] and hexagonal prism shaped cells are used in [5]. It is shown in [2] that the required number of nodes to monitor a 3D space is 43.25% fewer when the shape of the cell is truncated octahedron than the case where the cell is represented as either hexagonal prism or rhombic dodecahedron. In this paper we extend the work of [2] in the arena of wireless sensor networks where placing individual nodes exactly in any predetermined position is not possible. However, we assume that the sensor nodes are densely and uniformly distributed in a 3D space. In this paper, we exploit that redundancy to compensate the placement restriction. As opposed to [2], where a node is always placed at the center of a cell, here placement restriction requires that the active node can be located anywhere inside a cell. As a result the required sensing range is the diameter of a cell whereas in [2], the sensing range is the radius of the cell. Although this paper also finds that truncated octahedron shaped cells are the best, relative superiority of truncated octahedron model over cube, hexagonal prism and rhombic dodecahedron model is different from the values obtained in the context of the problem in [2]. So without any careful analysis, it should not be assumed that truncated octahedron shaped cells are best for any problem in 3D.

## 3. PROBLEM STATEMENT

The main assumptions and the goals of this work are defined as follows:

### *3.1 Assumptions*

- The sensors are uniformly distributed over a 3D space.
- All sensor nodes are identical. For example, they have identical transmission range $r_t$ and identical energy source (battery). Transmission is omni-directional and the transmission region of each node can be represented by a sphere of radius $r_t$, having the node at its center.
- The transmission range $r_t$ is much smaller than the length, the width, or the height of the 3D space to be covered, so that the boundary effect is negligible and hence can be ignored.
- There is a localization component in each sensor node that allows it to determine its location in the 3D space.

### 3.2 *Goals*

- Given any fixed $r_t$, find the best way to partition the network into cells in three-dimension such that the number of active nodes (i.e., number of cells) in the whole network is minimized while maintaining full connectivity.
- Find the minimum sensing range in terms of transmission radius such that any point in the network can be sensed by at least one active node.
- Find an algorithm so that such a partition (i.e., each sensor node knows in which cell they belong) can be made in a fast, efficient and distributed manner.

## 4. ANALYSIS

We partition the 3D space into equal sized non-overlapping cells and keep just one sensor active inside each cell. In order to maintain connectivity, the distance between an active node and any of its first-tier neighboring active nodes can not exceed the transmission range. Since the active node of a cell can reside anywhere inside a cell, the maximum distance between a point in a cell and a point in any of its first-tier neighbors can not exceed the transmission range $r_t$. Clearly, the shape of the cell must be a polyhedron that tessellates a 3D space. Among the polyhedrons that tessellate a 3D space the following four are most prominent and worthy of consideration: cube, rhombic dodecahedron, hexagonal prism and truncated octahedron (See [2] for details). For a fixed transmission range $r_t$, we first find out the maximum radius of a cell for different shapes of the cell.

### 4.1 *Maximum Radius of a Cell*

In this section we calculate the maximum radius of a cell when the shape of the cell is cube (*CB*), rhombic dodecahedron (*RD*), hexagonal prism (*HP*) and truncated octahedron (*TO*), given a fixed transmission radius $r_t$.

#### 4.1.1 Cube

If a cell has the shape of a cube, then the number of first tier neighboring cells is 26. In this case, there are three types of neighboring cells (See Figure 1)

1. Neighboring cells share a common plane (6 such neighbors) (*Type* 1)
2. Neighboring cells share a common line (12 such neighbors) (*Type* 2)
3. Neighboring cells share just a common point (8 such neighbors) (*Type* 3)

Suppose that the radius of a cube is $R$. Then the largest distance between any two points of *Type* 1 neighboring cells is $R2\sqrt{2}$ =2.828427$R$; for *Type* 2 and *Type* 3 neighbors, it is $R2\sqrt{3}$ =3.4641$R$ and 4$R$, respectively. So the active node of a cell can communicate with active nodes of all first-tier neighboring cells if the maximum radius of the Cube shaped cell is $r = \frac{r_t}{\max\left(2\sqrt{2}, 2\sqrt{3}, 4\right)} = \frac{r_t}{4} = 0.25 r_t$

### 4.1.2 Hexagonal Prism

If a cell has the shape of a hexagonal prism (*HP*), then the number of first tier neighboring cells is 20. In this case, there are three types of neighboring cells (See Figure 2 )

1. Neighboring cells share a common square plane (6 such neighbors) (*Type* 1)
2. Neighboring cells share a common hexagonal plane (2 such neighbors) (*Type* 2)
3. Neighboring cells share a common line (12 such neighbors) (*Type 3*)

Suppose that each side of a hexagonal face of *HP* is of length *a*, and the height of *HP* is *h.* In a *HP* with optimal height, $h = a\sqrt{2}$ [2]. So the radius of *HP* is $R = \sqrt{a^2 + \frac{a^2}{2}} = a\sqrt{\frac{3}{2}}$.

So maximum distance between any two points of *Type* 1 neighbors is

$$\sqrt{\left(a\sqrt{13}\right)^2 + h^2} = \sqrt{\left(a\sqrt{13}\right)^2 + (a\sqrt{2})^2} = a\sqrt{15} = R\sqrt{\frac{2\times 15}{3}} = R\sqrt{10} = 3.16227766\,\mathrm{R}$$

and, maximum distance between any two points of *Type* 2 neighbors is

$$\sqrt{(2a)^2 + (2h)^2} = \sqrt{(2a)^2 + (2a\sqrt{2})^2} = a\sqrt{12} = R\sqrt{\frac{2\times 12}{3}} = R\sqrt{8} = 2.828427R$$

Finally, maximum distance between any two points of *Type* 3 neighbors is

$$\sqrt{\left(a\sqrt{13}\right)^2 + (2h)^2} = \sqrt{\left(a\sqrt{13}\right)^2 + (a2\sqrt{2})^2} = a\sqrt{21} = R\sqrt{\frac{2\times 21}{3}} = R\sqrt{14} = 3.741657387R$$

So the active node of a cell can communicate with active nodes of all neighboring cells if the maximum radius of the *HP* shaped cell is $r = \frac{r_t}{\max\left(\sqrt{10}, \sqrt{8}, \sqrt{14}\right)} = \frac{r_t}{\sqrt{14}} = 0.26726 r_t$

### 4.1.3 Rhombic Dodecahedron

If a cell has the shape of a rhombic dodecahedron (*RD*), then the number of first tier neighboring cells is 18. In this case, there are two types of neighboring cells (See Figure 3 )

1. Neighboring cells share just a common point (6 such neighbors) (*Type* 1)
2. Neighboring cells share a common plane (12 such neighbors) (*Type* 2)

Suppose that the radius of a *RD* cell is *R*. Then, maximum distance between any two points of *Type* 1 neighbors is 4*R.* It is very difficult to analytically find the maximum distance between any two points of two *Type* 2 neighbors. So we find this distance by exhaustive search. We calculate the maximum distance between any two points of *Type* 2 neighbors by calculating the maximum distance between any two vertices of two *Type 2* neighbors. If the center of the *RD* has the coordinate ($x_c, y_c, z_c$), then the coordinates of its 14 vertices can be as follows (note that other values are also possible based on the orientation of *RD*, however that does not change the ultimate result):

$\left(x_c + R/\sqrt{2}, y_c, z_c + R/2\right)$, $\left(x_c + R/\sqrt{2}, y_c, z_c - R/2\right)$, $\left(x_c + R/\sqrt{2}, y_c + R/\sqrt{2}, z_c\right)$, $\left(x_c + R/\sqrt{2}, y_c - R/\sqrt{2}, z_c\right)$, $\left(x_c - R/\sqrt{2}, y_c, z_c + R/2\right)$, $\left(x_c - R/\sqrt{2}, y_c, z_c - R/2\right)$, $\left(x_c - R/\sqrt{2}, y_c + R/\sqrt{2}, z_c\right)$, $\left(x_c - R/\sqrt{2}, y_c - R/\sqrt{2}, z_c\right)$, $\left(x_c, y_c + R/\sqrt{2}, z_c + R/2\right)$, $\left(x_c, y_c + R/\sqrt{2}, z_c - R/2\right)$, $\left(x_c, y_c - R/\sqrt{2}, z_c + R/2\right)$, $\left(x_c, y_c - R/\sqrt{2}, z_c - R/2\right)$, $\left(x_c, y_c, z_c + R\right)$, $\left(x_c, y_c, z_c - R\right)$.

Suppose that the coordinate of information sink (*IS*) is (*cx,cy,cz*). Then in *RD* tessellation of 3D space, the coordinate of the center of each of the *RD* cell is given by $\left(cx + (2u + w)\frac{R}{\sqrt{2}}, cy + (2v + w)\frac{R}{\sqrt{2}}, cz + wR\right)$, where *u*, *v* and *w* are integer coordinates [2]. To calculate the maximum distance between any two points of two neighboring rhombic dodecahedron, assume without loss of generality that *cx*=*cy*=*cz*=0; *w*=*v*=0; *u*=0 and 1. Then the centers of two *Type* 2 neighboring *RD*s are (0,0,0) and ($R\sqrt{2}$,0,0). We then calculate the maximum distance of any of the 14 vertices of the *RD* centered at (0,0,0) with any of the 14 vertices of the *RD* centered at ($R\sqrt{2}$,0,0) and find that the maximum distance is between the following pair of vertices having coordinates $\left(-R/\sqrt{2}, R/\sqrt{2}, 0\right)$ and $\left(R\sqrt{2} + R/\sqrt{2}, -R/\sqrt{2}, 0\right)$; $\left(-R/\sqrt{2}, -R/\sqrt{2}, 0\right)$ and $\left(R\sqrt{2} + R/\sqrt{2}, R/\sqrt{2}, 0\right)$; which is $R\sqrt{10}$ =3.16227766*R*.

So the active node of a cell can communicate with active nodes of all first-tier neighboring cells if the maximum radius of the *RD* shaped cell is $r = \frac{r_t}{\max\left(4, \sqrt{10}\right)} = \frac{r_t}{4} = 0.25r_t$

### 4.1.4 Truncated Octahedron

If a cell has the shape of a Truncated Octahedron (*TO*), then the number of first tier neighboring cells is 14. In this case, there are two types of neighboring cells (See Figure 4 )

1. Neighboring cells share a common square plane (6 such neighbors) (*Type* 1)
2. Neighboring cells share a common hexagonal plane (8 such neighbors) (*Type* 2)

For both types of neighbors, we employ the same technique that is used for *Type* 2 neighboring cells of *RD*. A truncated octahedron has 8 regular hexagonal faces, 6 regular square faces, 24 vertices and 36 edges. If the center of a *TO* has the coordinate ($x_c$, $y_c$, $z_c$), then the coordinates of its 24 vertices can be as follows:
($x_c$-*d*, $y_c$+*d*/2, $z_c$), ($x_c$-*d*, $y_c$-*d*/2, $z_c$), ($x_c$-*d*, $y_c$, $z_c$+*d*/2), ($x_c$-*d*, $y_c$, $z_c$-*d*/2), ($x_c$-*d*/2, $y_c$+*d*, $z_c$), ($x_c$-*d*/2, $y_c$-*d*, $z_c$), ($x_c$-*d*/2, $y_c$, $z_c$+*d*), ($x_c$-*d*/2, $y_c$, $z_c$-*d*), ($x_c$, $y_c$+*d*, $z_c$+*d*/2), ($x_c$, $y_c$-*d*, $z_c$+*d*/2), ($x_c$, $y_c$+*d*/2, $z_c$+*d*), ($x_c$, $y_c$+*d*/2, $z_c$-*d*), ($x_c$, $y_c$+*d*, $z_c$-*d*/2), ($x_c$, $y_c$-*d*, $z_c$-*d*/2), ($x_c$, $y_c$-*d*/2, $z_c$+*d*), ($x_c$, $y_c$-*d*/2, $z_c$-*d*), ($x_c$+*d*/2, $y_c$+*d*, $z_c$), ($x_c$+*d*/2, $y_c$-*d*, $z_c$), ($x_c$+*d*/2, $y_c$, $z_c$+*d*), ($x_c$+*d*/2, $y_c$, $z_c$-*d*), ($x_c$+*d*, $y_c$+*d*/2, $z_c$), ($x_c$+*d*, $y_c$-*d*/2, $z_c$), ($x_c$+*d*, $y_c$, $z_c$+*d*/2), ($x_c$+*d*, $y_c$, $z_c$-*d*/2) where $d = \frac{2R}{\sqrt{5}}$ and the radius of *TO* is *R*.

Suppose that the coordinate of information sink (*IS*) is (*cx,cy,cz*). Then in *TO* tessellation of 3D space, the coordinate of the center of each of the *TO* cell can be represented as $\left(cx+(2u+w)\frac{2R}{\sqrt{5}}, cy+(2v+w)\frac{2R}{\sqrt{5}}, cz+w\frac{2R}{\sqrt{5}}\right)$ $=(cx+(2u+w)d, cy+(2v+w)d, cz+wd)$ where *u*, *v* and *w* are integer coordinates [2]. In order to calculate the maximum distance between any two points of two *Type* 1 neighboring *TO*s assume without loss of generality, *cx*=*cy*=*cz*=0; *w*=*v*=0; *u*=0 and 1. Then center of two neighboring *TO*s are (0,0,0) and (2*d*,0,0).

Using a computer program, we found that the maximum distance is between the following pair of vertices having coordinates: (-*d*, *d*/2, 0) and (2*d*+*d*, -*d*/2, 0); (-*d*, -*d*/2, 0) and (2*d*+*d*, *d*/2, 0); (-*d*, 0, *d*/2) and (2*d*+*d*, 0, -*d*/2); (-*d*, 0, -*d*/2) and (2*d*+*d*, 0, *d*/2); which is $d\sqrt{17}=\frac{2R}{\sqrt{5}}\sqrt{17}$ =3.6878177829*R*.

*TO*s centered at coordinate (0,0,0) and (*d,d,d*) are *Type* 2 neighbors. Using a computer program, we found that the maximum distance is between the following pair of vertices having coordinates: (-*d*, -*d/2*,0) and (*d+d, d+d/2, d*); (-*d*, 0, -*d/2*) and (*d+d, d, d+d/2*); (-*d*/2, -*d*, 0) and (*d+d/2, d+d, d*); (-*d*/2, 0, -*d*) and (*d+d/2, d, d+d*); (0, -*d*, -*d*/2) and (*d, d+d, d+d*/2); (0, -*d*/2, 0-*d*) and (*d, d+d/2, d+d*). Then the maximum distance is $d\sqrt{14}=\frac{2R}{\sqrt{5}}\sqrt{14}$ =3.34664*R*

So the active node of a cell can communicate with active nodes of all neighboring cells if the maximum radius of the *TO* shaped cell is $r=\frac{r_t}{\max\left(\frac{2\sqrt{17}}{\sqrt{5}},\frac{2\sqrt{14}}{\sqrt{5}}\right)}=\frac{r_t\sqrt{5}}{2\sqrt{17}}=0.271163r_t$

The results are summarized in Table I.

### 4.2 Minimum Sensing Range

Since an active node can be located anywhere inside a cell and still it must be able to sense any point inside the cell, the sensing range must be at least equal to the maximum distance between any two points of a cell. This maximum distance is essentially the diameter of a cell and equal to twice of the corresponding radius. So minimum sensing range of *CB*, *HP*, *RD* and *TO* is $2\times\frac{r_t}{4}=0.5r_t$, $2\times\frac{r_t}{\sqrt{14}}=0.53452r_t$, $2\times\frac{r_t}{4}=0.5r_t$ and $2\times\frac{r_t\sqrt{5}}{2\sqrt{17}}=0.542326r_t$, respectively (See Figure 5 ).

### 4.3 Sensors Determining Respective Cell ID

Using the value of maximum radius of a cell, we can calculate the coordinate of the centers of each of the cells according to the methods shown in [2]. Here we show how a sensor node can determine its cell id when the shape of the cell is *TO*. It is a simple exercise to determine the cell id under *CB, HP* and *RD* models using similar techniques.

For $TO$ model, we have $R = \frac{r_t\sqrt{5}}{2\sqrt{17}}$. From [2], we know the coordinate of the center of each cell is $\left(cx + (2u+w)\frac{2R}{\sqrt{5}}, cy + (2v+w)\frac{2R}{\sqrt{5}}, cz + w\frac{2R}{\sqrt{5}}\right)$ where $(cx, cy, cz)$ is the coordinate of the information sink (*IS*). So the coordinates of the centers of $TO$ shaped cells can be expressed by the generalized equation $\left(cx + (2u+w)\frac{r_t}{\sqrt{17}}, cy + (2v+w)\frac{r_t}{\sqrt{17}}, cz + w\frac{r_t}{\sqrt{17}}\right)$ where $(u,v,w)$ represents a unique cell by the integer coordinates.

Now the question is how a sensor can determine in which cell it belongs to. Assume that the node can determine its location and without loss of generality assume that its coordinate is $(x_s, y_s, z_s)$ and it also knows that the coordinate of the *IS* is $(cx, cy, cz)$. It wants to know its cell id $(u_s, v_s, w_s)$. The obvious brute force method is to check all possible values of $(u_s, v_s, w_s)$ and choose the cell whose center is has minimum Euclidean distance from the node. In other word,

$$(u_s, v_s, w_s) = \arg\min_{\substack{u\in\mathbf{Z},\\ v\in\mathbf{Z},\\ w\in\mathbf{Z}}} \left(x_s - cx - (2u+w)\frac{r_t}{\sqrt{17}}\right)^2 + \left(y_s - cy - (2v+w)\frac{r_t}{\sqrt{17}}\right)^2 + \left(z_s - cz - w\frac{r_t}{\sqrt{17}}\right)^2,$$

where $\mathbf{Z}$ is set of all integers.

However, we do not need to do exhaustive search. Since the value of a square term is never negative, we can set the value of the square terms to zero to get the values of $u_s, v_s$, and $w_s$. We have, $w_s = (z_s - cz)\sqrt{17}/r_t$. However, this value can be a fraction but we need integral values. We can get two possible integral values by taking ceiling (denoted by subscript $h$) and floor (subscript $l$):

$$u_l = \left\lfloor \frac{1}{2}\left((x_s - cx)\frac{\sqrt{17}}{r_t} - w_s\right)\right\rfloor \quad => \quad u_l = \left\lfloor \frac{1}{2}\left((x_s - cx)\frac{\sqrt{17}}{r_t} - (z_s - cz)\frac{\sqrt{17}}{r_t}\right)\right\rfloor = \left\lfloor (x_s - cx - z_s + cz)\frac{\sqrt{17}}{2r_t}\right\rfloor$$

$$u_h = \left\lceil \frac{1}{2}\left((x_s - cx)\frac{\sqrt{17}}{r_t} - w_s\right)\right\rceil \quad => \quad u_h = \left\lceil \frac{1}{2}\left((x_s - cx)\frac{\sqrt{17}}{r_t} - (z_s - cz)\frac{\sqrt{17}}{r_t}\right)\right\rceil = \left\lceil (x_s - cx - z_s + cz)\frac{\sqrt{17}}{2r_t}\right\rceil$$

$$v_l = \left\lfloor \frac{1}{2}\left((y_s - cy)\frac{\sqrt{17}}{r_t} - w_s\right)\right\rfloor \quad => \quad v_l = \left\lfloor \frac{1}{2}\left((y_s - cy)\frac{\sqrt{17}}{r_t} - (z_s - cz)\frac{\sqrt{17}}{r_t}\right)\right\rfloor = \left\lfloor (y_s - cy - z_s + cz)\frac{\sqrt{17}}{2r_t}\right\rfloor$$

$$v_h = \left\lceil \frac{1}{2}\left((y_s - cy)\frac{\sqrt{17}}{r_t} - w_s\right)\right\rceil \quad => \quad v_h = \left\lceil \frac{1}{2}\left((y_s - cy)\frac{\sqrt{17}}{r_t} - (z_s - cz)\frac{\sqrt{17}}{r_t}\right)\right\rceil = \left\lceil (y_s - cy - z_s + cz)\frac{\sqrt{17}}{2r_t}\right\rceil$$

$$w_l = \left\lfloor (z_s - cz)\frac{\sqrt{17}}{r_t}\right\rfloor; \qquad w_h = \left\lceil (z_s - cz)\frac{\sqrt{17}}{r_t}\right\rceil \tag{1}$$

Thus we have eight possible values of $(u_s, v_s, w_s)$. Each node has to calculate its distance from each of the eight centers and choose the minimum one i.e.,

$$(u_s, v_s, w_s) = \arg \min_{\substack{u \in \{u_l, u_h\}, \\ v \in \{v_l, v_h\}, \\ w \in \{w_l, w_h\}}} \sqrt{\begin{array}{l} \left( x_s - cx - (2u+w)\dfrac{r_t}{\sqrt{17}} \right)^2 + \\ \left( y_s - cy - (2v+w)\dfrac{r_t}{\sqrt{17}} \right)^2 + \left( z_s - cz - w\dfrac{r_t}{\sqrt{17}} \right)^2 \end{array}} \qquad (2)$$

The complete topology control algorithm is provided in Algorithm 1. Instead of calculating the distance from each of the eight centers, we could simply take the nearest integer value for $u_s$,$v_s$, $w_s$. But as shown in Section 5, this approximation leads to incorrect prediction of cell id in almost one quarter of the cases.

**Algorithm 1: Sensor Topology Control Algorithm**

1. *IS* broadcast a message to all nodes that contains its location $(cx,cy,cz)$.
2. Once a sensor node knows the location of *IS* and transmission radius $r_t$ and calculates its own location $(x_s,y_s,z_s)$ using any localization or other techniques, it can determine in which cell it belongs to in the following way. The sensor node calculates six values $u_l, v_l, w_l, u_h, v_h, w_h$ according to (1). Using the combination of these values eight cells can be identified, i.e., $(u_l, v_l, w_l)$, $(u_l, v_l, w_h)$, $(u_l, v_h, w_l)$, $(u_l, v_h, w_h)$, $(u_h, v_l, w_l)$, $(u_h, v_l, w_h)$, $(u_h, v_h, w_l)$, $(u_h, v_h, w_h)$. The sensor then just chooses the cell whose center has the least Euclidean distance from it as shown in (2).
3. Once a sensor find the value of $(u_s,v_s,w_s)$ that minimizes above distance, it identify itself with this cell id.

Since there are just eight possible combinations, above calculation for finding cell id involves a small constant number of local arithmetic operations. As cell id is a straightforward function of the location of a sensor, if a sensor knows the location of another sensor, it can readily calculate the cell id of that sensor.

Once sensors have their cell id, then sensors with same cell id can use any standard leader selection algorithms [11] to choose a leader among them which can act as the active node of that cell. All nodes that have same cell id are within the communication range of each other and the mechanism of keeping one node active among all the sensors with same cell id is essentially same for both 2D and 3D networks. Since the main focus of this paper is problems that are unique to 3D networks, we choose not to explore the issues that have already been studied in the context of 2D networks.

### *4.4 Number of Active Nodes and Network Lifetime*

Since at a time the number of active nodes in a cell is one, total number of active nodes in a network is equal to the number of cells in the network. Since according to the assumption the boundary effect is negligible and can be ignored, the number of cells in a network is inversely proportional to the volume of the network. So the number of active nodes for different shapes of the cell can be easily calculated if the volume of the cell is known.

The volume of a cube (*CB*), hexagonal prism (*HP*), rhombic dodecahedron (*RD*) and truncated octahedron (*TO*) cell of radius $R$ is $\frac{8R^3}{3\sqrt{3}} = 1.5396R^3$, $2R^3$, $2R^3$ and $\frac{32R^3}{5\sqrt{5}} = 2.862R^3$, respectively.

Using the maximum radius for various shapes from, we have the volume of a cell for various shapes:

For *CB*: $V_{rc}^{CB} = 8\left(\frac{r_t}{4}\right)^3 \Big/ 3\sqrt{3} = \frac{r_t^3}{24\sqrt{3}} = 0.024056 r_t^3$; For *HP*: $V_{rc}^{HP} = 2\left(\frac{r_t}{\sqrt{14}}\right)^3 = \frac{r_t^3}{7\sqrt{14}} = 0.03818 r_t^3$

For *RD*: $V_{rc}^{RD} = 2\left(\frac{r_t}{4}\right)^3 = \frac{r_t^3}{32} = 0.03125 r_t^3$; For *TO*: $V_{rc}^{TO} = 32\left(\frac{r_t\sqrt{5}}{2\sqrt{17}}\right)^3 \Big/ 5\sqrt{5} = \frac{4}{17\sqrt{17}} r_t^3 = 0.057 r_t^3$

So the active nodes required by *CB*, *HP* and *RD* model is, respectively, $\frac{96\sqrt{3}}{17\sqrt{17}} = 2.372239$, $\frac{28\sqrt{14}}{17\sqrt{17}} = 1.49468$ and $\frac{128}{17\sqrt{17}} = 1.82615$ times of the active nodes required by *TO* model (See Figure 7 ).

Now, we use a simplified model to calculate the network lifetime for different shapes of a cell. Since transmission radius is same for all cases, it can be assumed that a node consumes same amount of power for transmission for all different shapes. If we ignore the power consumption discrepancy due to difference in the number of packets relayed by a node, then the lifetime of an individual node is roughly same in all cases. So lifetime of a cell is proportional to the number of nodes in a cell. Since the assumption is that the sensor nodes are uniformly distributed, the number of nodes in a cell is proportional to the volume of the cell. So in general, we can say that the ratio of network lifetime in different models is essentially the ratio of volume of a cell under those models. Then network lifetime of CB model is $\frac{17\sqrt{17}}{96\sqrt{3}} = 0.42154 = 42.154\%$ of that of *TO* model. This value is $\frac{17\sqrt{17}}{28\sqrt{14}}$ =0.669=66.9% for *HP* model and $\frac{17\sqrt{17}}{128}$ =0.5476=54.76% for *RD* model. Table II and Figure 6 summarize the results.

## 5. SIMULATION

We use simulation to validate that each sensor node can determine their respective cell id correctly according to (1) and (2). We do it by randomly generating the location of a sensor, then calculating its cell id according to (1) and (2) and also determining its cell id by exhaustive search. In a very large number of trials, we found that in every case our equations (ceil and floor approach) can predict the cell id correctly. However, further effort to simplifying the prediction process does not work. For example, instead of determining eight possible cell ids and calculating distance with each of them, we could find one cell id directly by taking the nearest integer value of *u, v, w* coordinates. Our simulation shows that this approach makes false predictions in roughly one quarter of the cases (See Table III and Figure 8 ).

## 6. DISCUSSIONS AND FUTURE WORK

Active nodes can use their cell id as their address. A greedy geographic routing scheme can work here as follows: source active node writes its cell id and destination active node's cell id in the packet. Suppose that the source cell id is $(u_s, v_s, w_s)$ and destination cell id is $(u_d, v_d, w_d)$. Then the source sends this packet to a neighbor with cell id $(u_i, v_i, w_i)$ such that $(u_d - u_i)^2 + (v_d - v_i)^2 + (w_d - w_i)^2 < (u_d - u_s)^2 + (v_d - v_s)^2 + (w_d - w_s)^2$. Then the active node with cell id $(u_i, v_i, w_i)$ sends this packet to a neighbor with cell id $(u_j, v_j, w_j)$ such that

$$(u_d - u_j)^2 + (v_d - v_j)^2 + (w_d - w_j)^2 < (u_d - u_i)^2 + (v_d - v_i)^2 + (w_d - w_i)^2 .$$

If more than one neighbor satisfies above criteria (most often which is actually the case), then the least loaded active node, the active node with the highest energy or just randomly one of them can be chosen. When the shape of each cell is truncated octahedron, each cell has 14 neighboring cells. The neighboring cells of a cell having cell id $(u_1, v_1, w_1)$ have the following ids: $(u_1+1, v_1, w_1)$, $(u_1-1, v_1, w_1)$; $(u_1, v_1+1, w_1)$, $(u_1, v_1-1, w_1)$; $(u_1-1, v_1-1, w_1+2)$, $(u_1+1, v_1+1, w_1-2)$; $(u_1, v_1, w_1+1)$, $(u_1, v_1, w_1-1)$; $(u_1-1, v_1, w_1+1)$, $(u_1+1, v_1, w_1-1)$; $(u_1, v_1-1, w_1+1)$, $(u_1, v_1+1, w_1-1)$; $(u_1-1, v_1-1, w_1+1)$, $(u_1+1, v_1+1, w_1-1)$. So it requires a small constant number of arithmetic operations to choose the optimal neighboring node to forward a packet.

Above simple approach works well when all active nodes are always connected with all of their neighboring active nodes. However, this greedy scheme might not work in all possible scenarios. In the presence of obstruction, there is a possibility that the packet reaches a dead end where no neighboring active node satisfy the criteria mentioned above and the packet is yet to reach the destination. In 2D, face routing can be deployed in such scenario [9] [6]. An extension of [9] in 3D is an interesting problem to look at. Localization in 3D space is also another interesting area of future research.

## 7. CONCLUSION

In this paper, we investigate the topology control and network lifetime issues in three-dimensional wireless sensor networks where sensor nodes are deployed in a 3D space unlike most current works that assume that the nodes are placed in a 2D plane. Our results show that partitioning the 3D space into truncated octahedron shaped cells such that distance between the furthest points of two first-tier neighbor cells equals the transmission range of the nodes and keeping only one node active at a time inside each cell minimizes the number of active nodes (thus maximizes network lifetime) while maintaining full connectivity. Full coverage can be achieved if the sensing range is at least 0.542326 times the transmission radius. If the requirement is such that any point in the 3D space has to be sensed or monitored by at least *k* sensors, then keeping *k* nodes active in each cell at a time fulfils that objective. We also provide a distributed topology control algorithm that allows a sensor node to determine its cell id using a few simple local arithmetic operations provided that the location information is available.

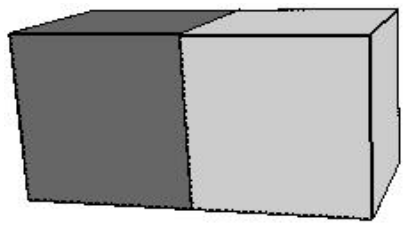

(a) *Type* 1 Neighbors

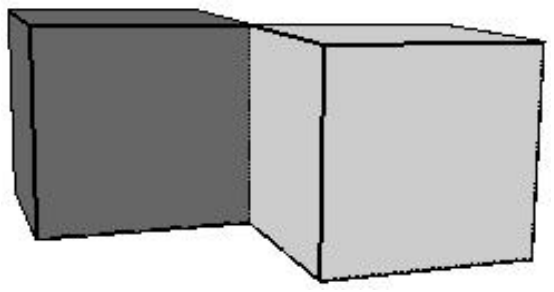

(b) *Type* 2 Neighbors

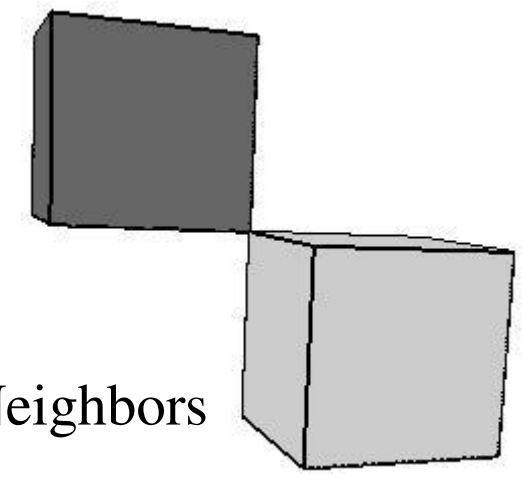

(c) *Type* 3 Neighbors

Figure 1 : Different types of neighbors in cube tessellation of 3D space

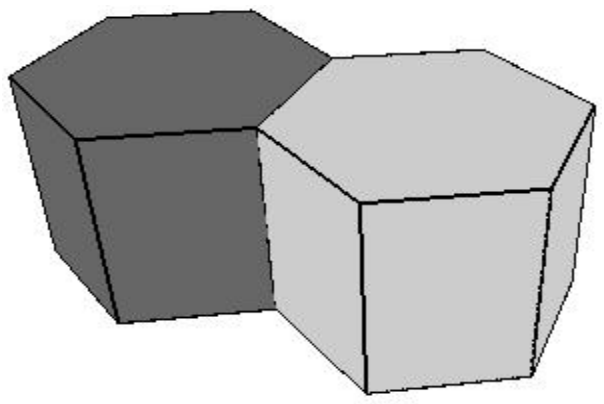

(a) *Type* 1 Neighbors

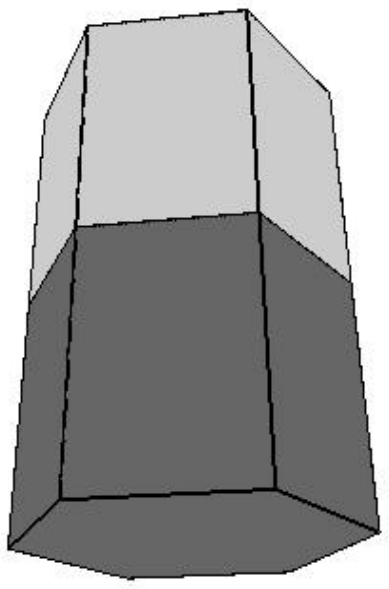

(b) *Type* 2 Neighbors

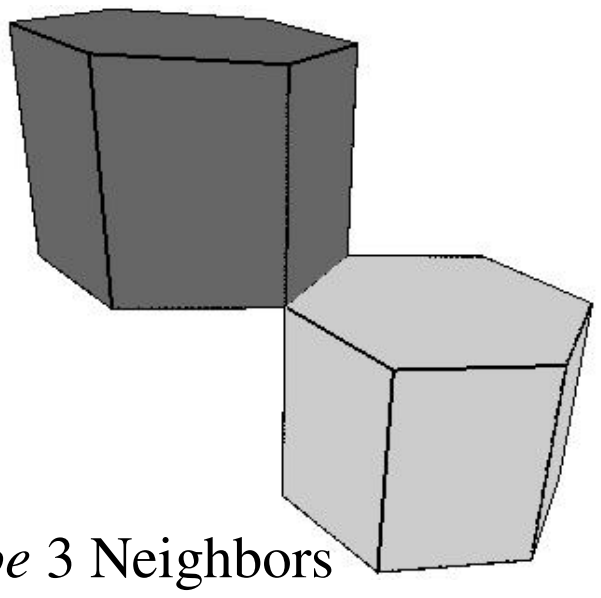

(c) *Type* 3 Neighbors

Figure 2 : Different types of neighbors in hexagonal prism tessellation of 3D space

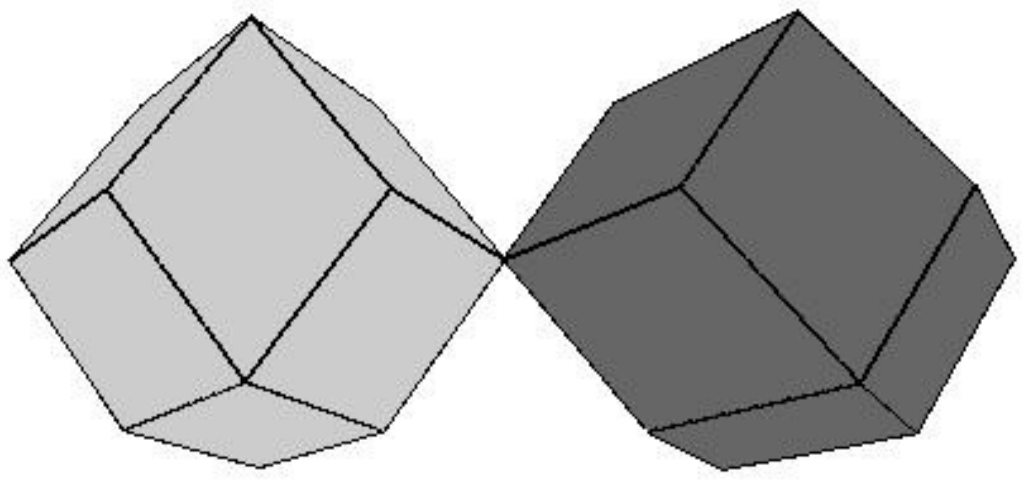

(a) *Type* 1 Neighbors

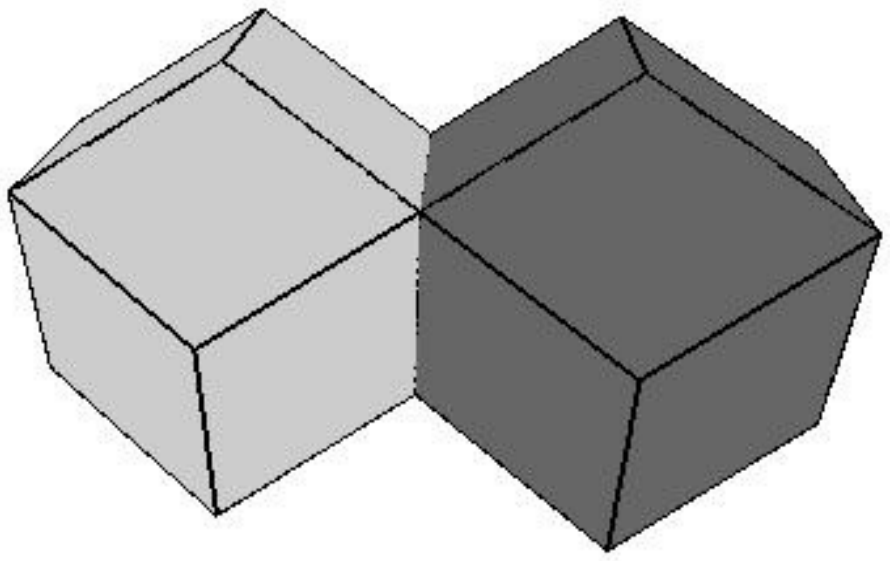

(b) *Type* 2 Neighbors

Figure 3 : Different types of neighbors in rhombic dodecahedron tessellation of 3D space

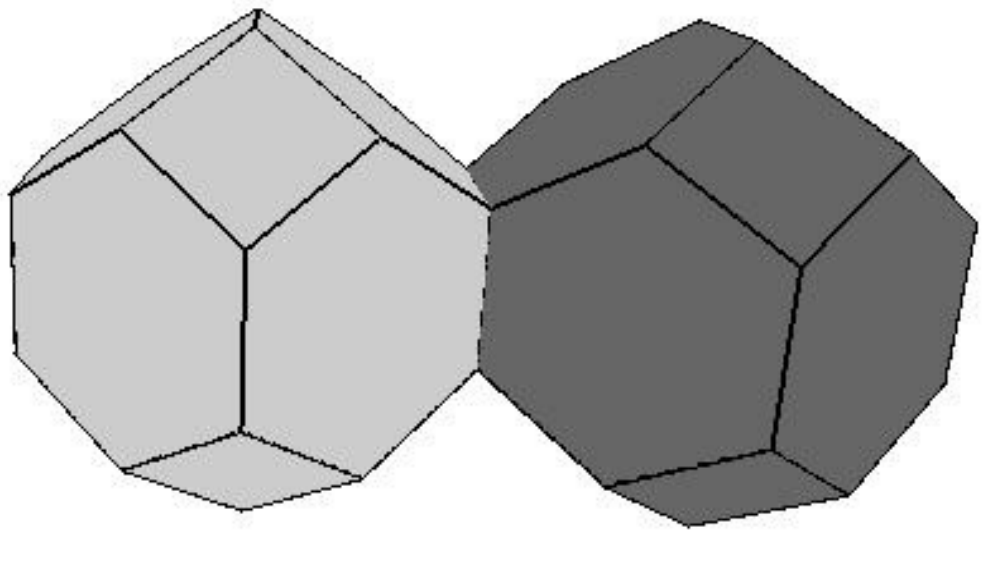

(a) *Type* 1 Neighbors

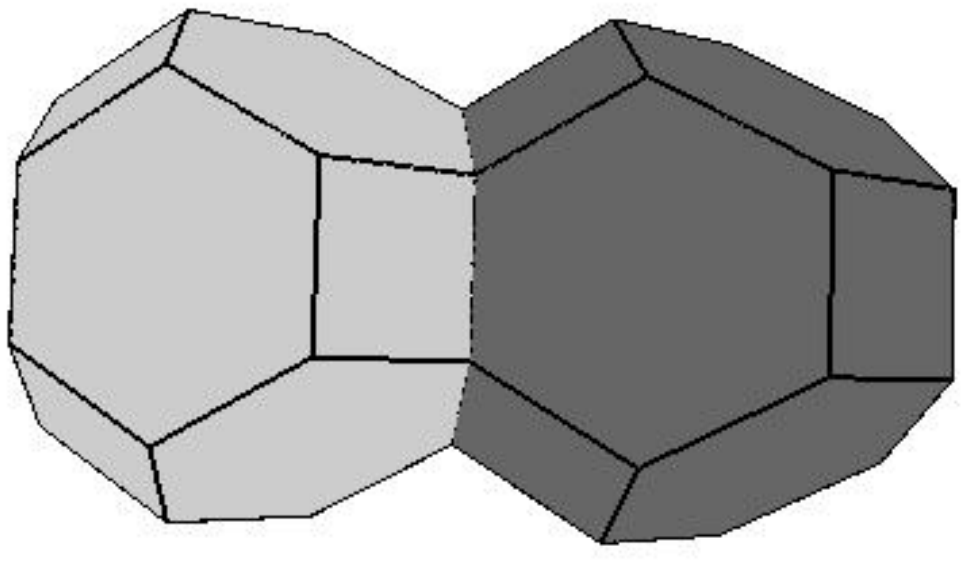

(b) *Type* 2 Neighbors

Figure 4 : Different types of neighbors in truncated octahedron tessellation of 3D space

**Table I: Maximum Radius of a Cell (also Minimum Sensing Range) for Various Shapes of the Cell**

| Shape of a Cell | Cube (*CB*) | Hexagonal Prism (Optimized Height) (*HP*) | Rhombic Dodecahedron (*RD*) | Truncated Octahedron (*TO*) |
|---|---|---|---|---|
| Number of first-tier Neighbors | 26 | 20 | 18 | 14 |
| Furthest distance between any two points of two neighboring partitions (i.e., transmission range $r_t$) | $4R$ (For two cube having a point in common)<br><br>$R2\sqrt{2}$ =2.828427$R$ (for two cube having a common face)<br><br>$R2\sqrt{3}$ =3.4641$R$ (for two cube having a line in common) | $R\sqrt{14}$ =3.741657387$R$ (for two HP having a shared line)<br><br>$R\sqrt{10}$ =3.16227766$R$ (for two HP having a shared square face)<br>$R\sqrt{8}$ =2.828427$R$ (for two HP having a shared hexagonal face) | $4R$ (for two RD having a single point in common)<br><br>$R\sqrt{10}$ =3.16227766$R$ (for two RD having a shared face) | $\frac{2R\sqrt{17}}{\sqrt{5}}$ =3.6878177829$R$ (for two TO sharing a square face)<br>$\frac{2R}{\sqrt{5}}\sqrt{14}$ =3.34664$R$ (for two TO sharing a hexagonal face) |
| Maximum radius of a cell for a fixed transmission range $r_{t,}$ | $R = \frac{r_t}{4} = 0.25 r_t$ | $R = \frac{r_t}{\sqrt{14}}$ $= 0.26726 r_t$ | $R = \frac{r_t}{4}$ $= 0.25 r_t$ | $R = \frac{r_t\sqrt{5}}{2\sqrt{17}}$ $= 0.271163 r_t$ |
| **Minimum Sensing range** | $0.5r_t$ | $0.53452r_t$ | $0.5r_t$ | $0.542326r_t$ |

**Table II: Number of Active Nodes and Network Lifetime**

| Model | Number of Active Nodes compare to *TO* model | Network Lifetime compare to *TO* model |
|---|---|---|
| Cube (*CB*) | 2.372239 | 42.154% |
| Hexagonal Prism (*HP*) | 1.49468 | 66.9% |
| Rhombic Dodecahedron (*RD*) | 1.82615 | 54.76% |
| Truncated Octahedron (*TO*) | 1 | 100% |

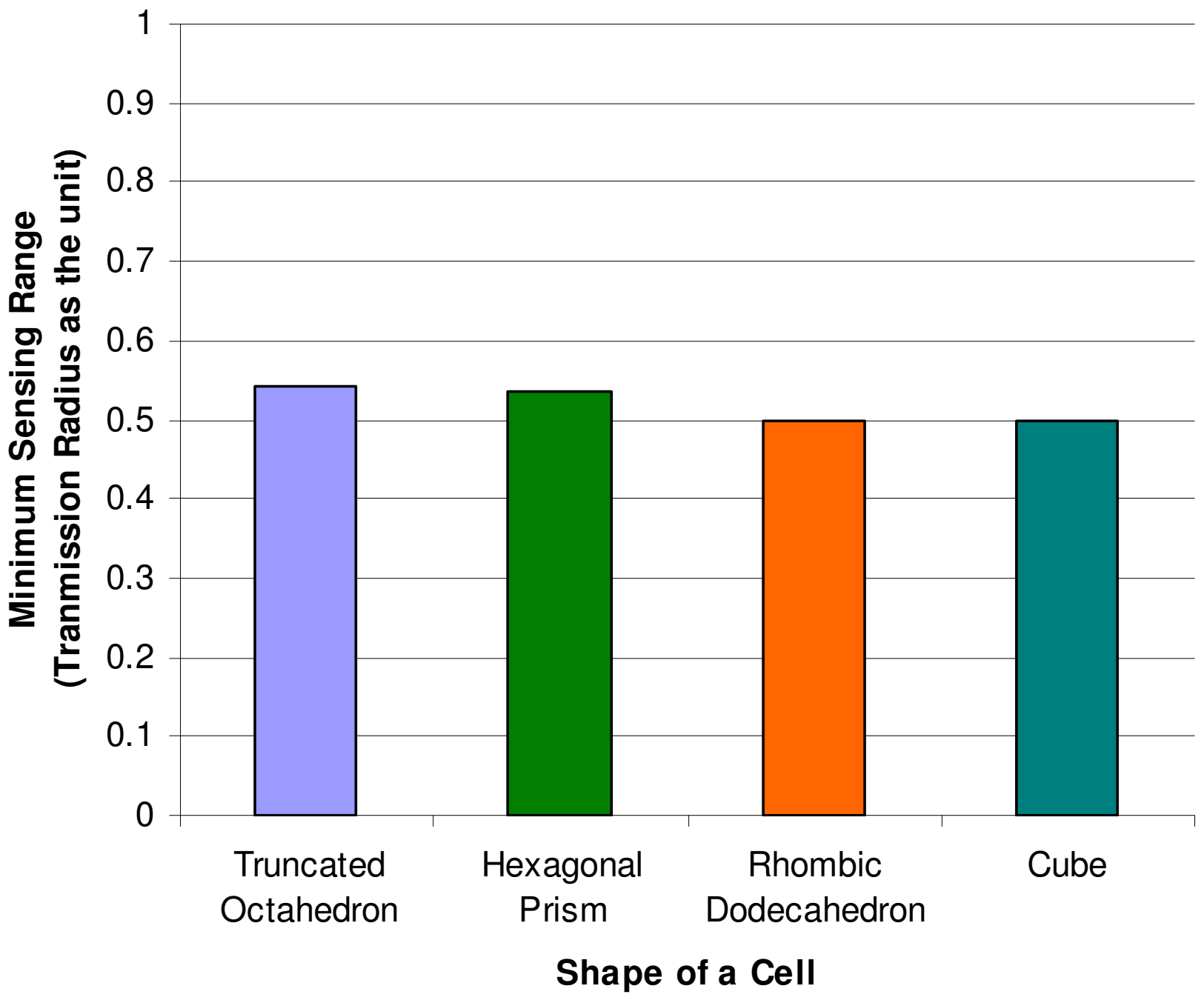

Figure 5 : Minimum Sensing Range

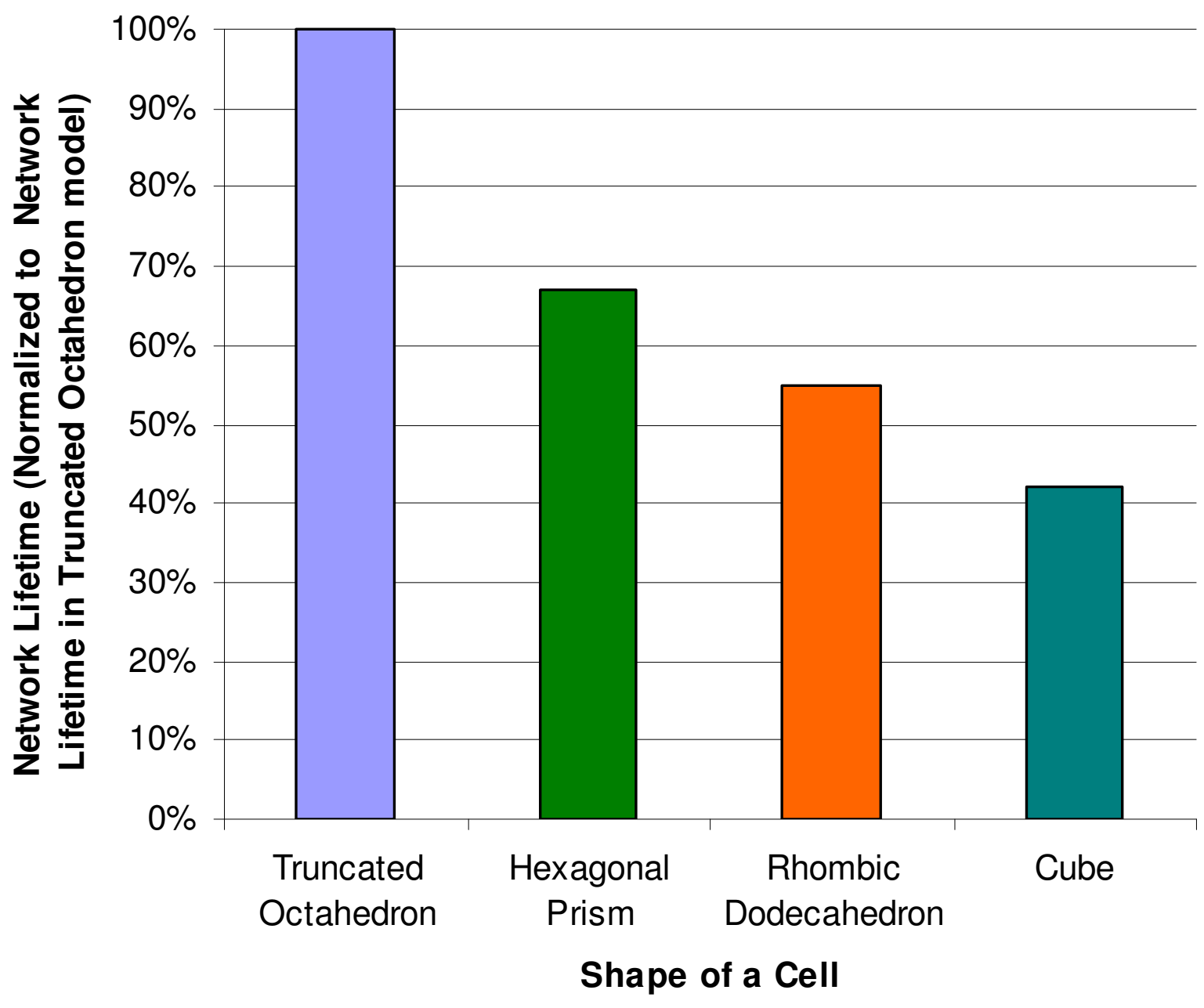

Figure 6 : Network Lifetime

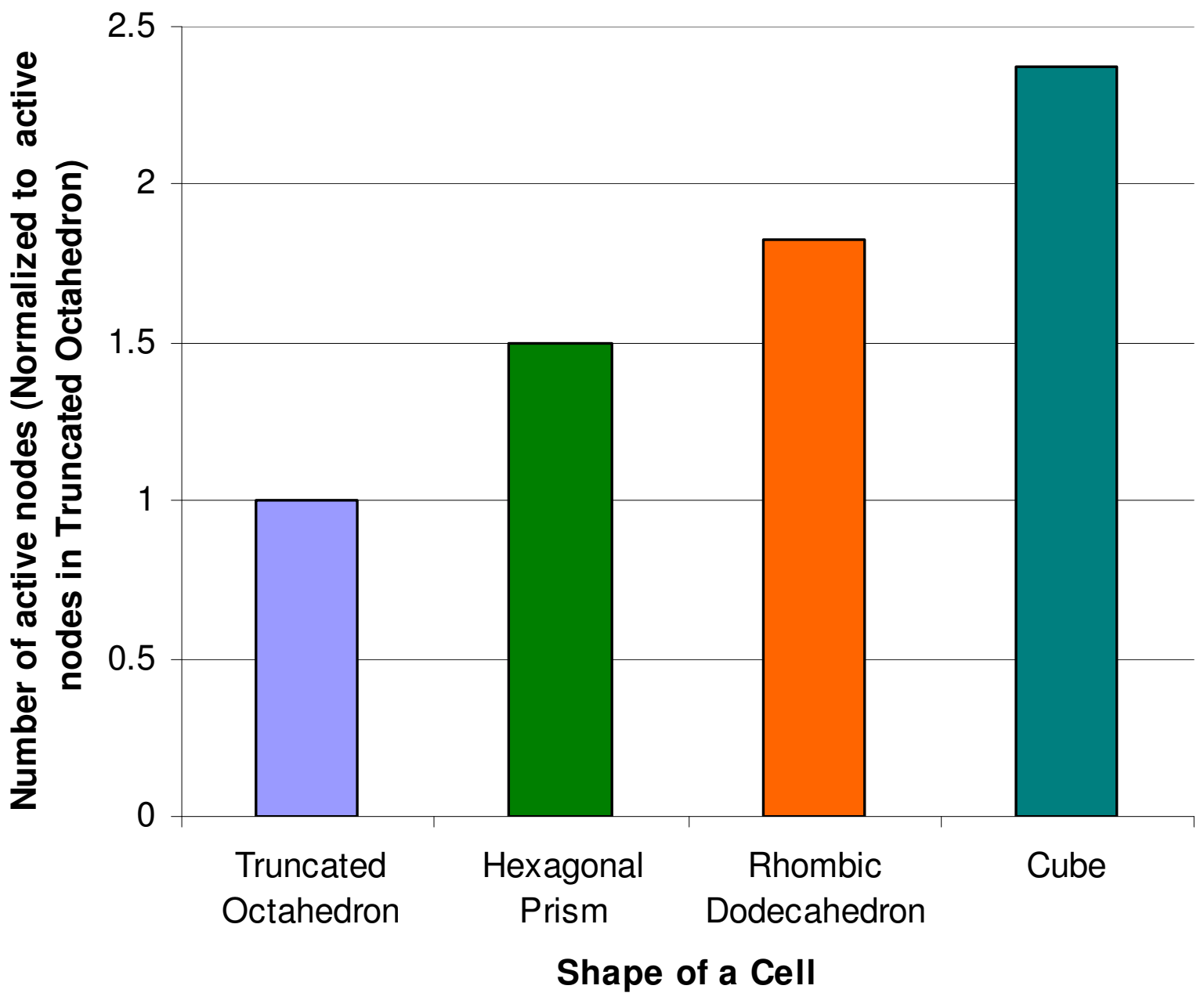

Figure 7 : Number of Active Nodes

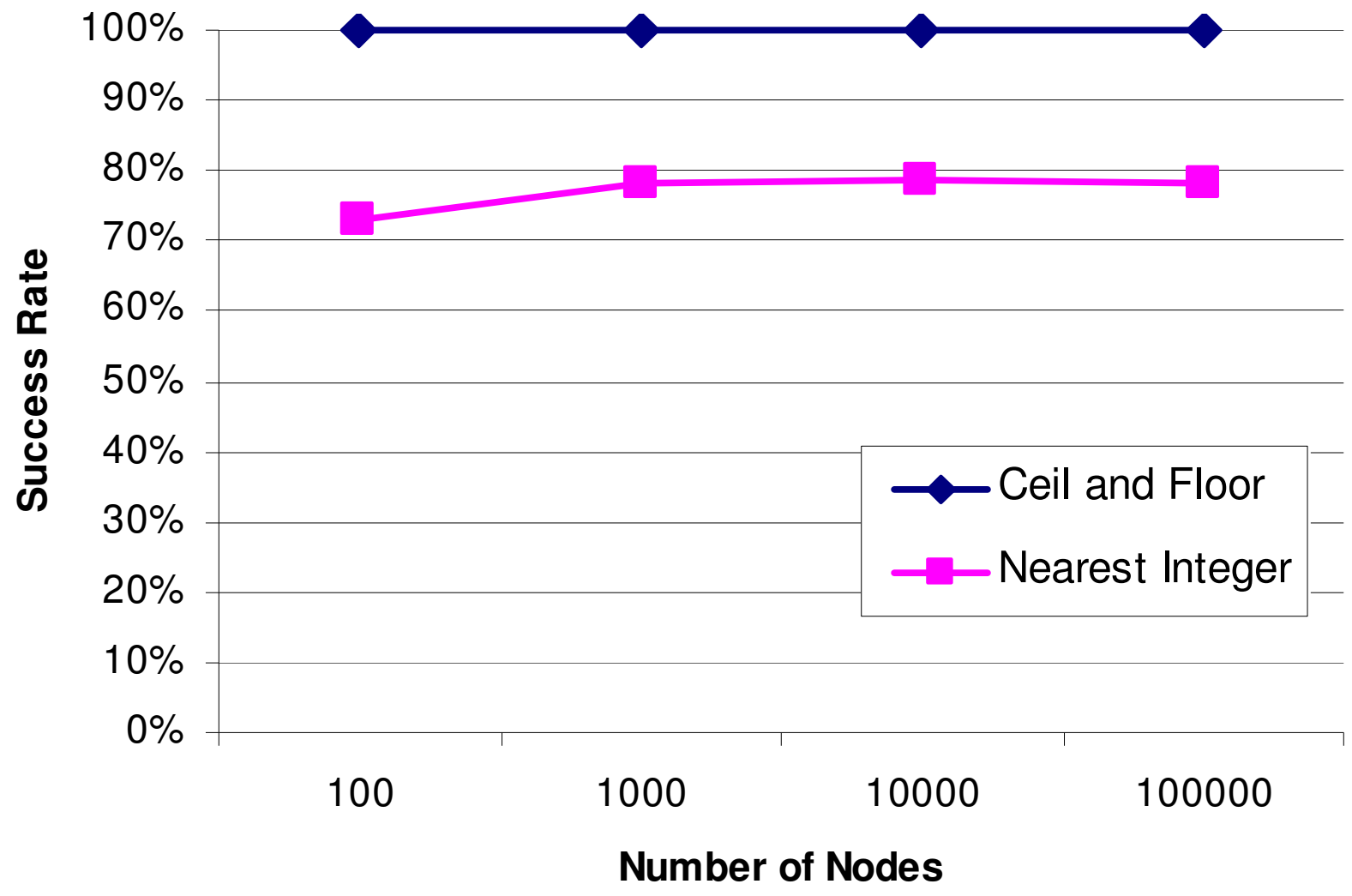

Figure 8 : Cell ID Prediction Accuracy

**Table III: Cell ID Prediction Accuracy**

| Number of Sensor Nodes | Total Correct Cell ID Prediction using (1) and (2) (Ceil and Floor) | Total Correct Cell ID Prediction Using Nearest Integer Approach |
|---|---|---|
| 100 | 100 | 73 |
| 1000 | 1000 | 782 |
| 10000 | 10000 | 7862 |
| 100000 | 100000 | 78354 |